# A real-time design based on FPGA for Expeditious Error Reconciliation in QKD system

Ke Cui, Jian Wang, *IEEE Member*, Hong-fei Zhang, Chun-li Luo, Ge Jin, Teng-yun Chen

*Abstract*—For high-speed quantum key distribution systems, error reconciliation is often the bottleneck affecting system performance. By exchanging common information through a public channel, the identical key can be generated on both communicating sides. However, the necessity to eliminate disclosed bits for security reasons lowers the final key rate. To improve this key rate, the amount of disclosed bits should be minimized. In addition, decreasing the time spent on error reconciliation also improves the key rate. In this paper we introduce a practical method for expeditious error reconciliation implemented in a Field Programmable Gate Array for a discrete variable quantum key distribution system, and illustrate the superiority of this method to other similar algorithms running on a PC. Experimental results demonstrate the rapidity of the proposed protocol.

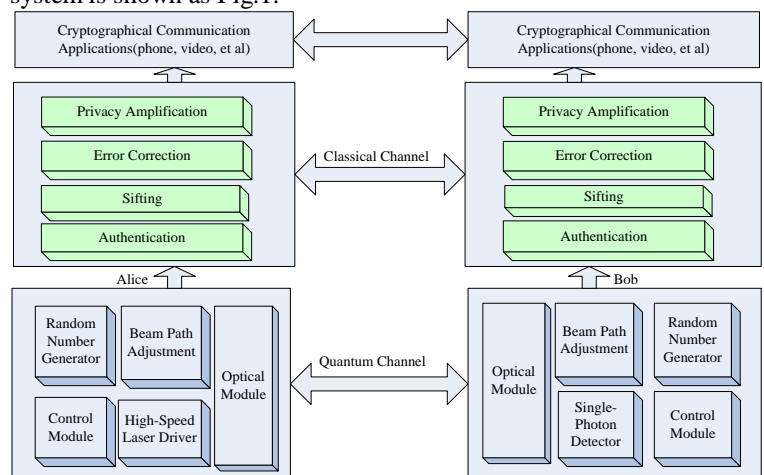

Fig.1 Typical structure of a QKD system

## I. INTRODUCTION

It has been more than 20 years since the introduction of quantum cryptography, which can be considered as the first practical achievement in quantum communication. Articles[1]-[3] give an overview of quantum key distribution(QKD). The goal of QKD is to create an absolutely secure key string between Alice and Bob (two sides of QKD system) and thus the key can be used in communication in one-time-pad manner. QKD generally contains the following three steps: (1) raw key sifting-Alice and Bob compare their basis of measurements and only the qubits having the same basis are retained. (2) error reconciliation-Alice and Bob correct errors in the sifted key by exchanging public information. This step can contain several passes. (3) privacy amplification-By applying universal hash functions[4] that map a long bit string to a shorter one, the information gained by the eavesdropper (Eve) can be made arbitrarily small. A typical structure of a QKD system is shown as Fig.1.

The second step in QKD, error reconciliation, is the focus of this work. Aside from any interference by Eve, errors inevitably exist due to imperfect equipment. These errors must be eliminated by exchanging information through a public channel. The first error reconciliation protocol was BBBSS[3]. Later on, other protocols like Cascade[5]-[7], AYHI[8], Winnow[9], the low-density parity-check(LDPC) protocol[10],[11] were used. The Cascade protocol was widely used in early QKD systems, because of its high efficiency for leaked information compared with the Shannon entropy limit. However, the Cascade protocol requires many interactions between the two sides resulting in degradation in speed. The Winnow protocol speeds the error correction process by adopting the Hamming code. The Hamming code corrects an error when only one error is in the code block within a single interaction, thus decreasing the total interaction frequency. The LDPC protocol is a one-way reconciliation method since only a single occurrence of information transfer is required for the entire key string. However, this protocol requires a LDPC decoder with high computational complexity. Recent work concerning LDPC developed some useful techniques[12] to facilitate its utility in QKD systems and an important work[13] gave the results of the implementations of LDPC code on high-end Graphic Processing Units(GPUs) to improve its reconciliation speed.

Up until now most error reconciliation has been completed using software on a PC. The rapidly growing demand for a high final key rate gives rise to a bottleneck in the entire QKD system.

Manuscript received March 13, 2012; revised July 13, 2012; accepted September 28, 2012. This work was supported by Chinese of Academy of Science, the National Fundamental Research Program of China under Grant 2006CB921900, the National High Technology Research and Development Program (863 Program) of China under Grant 2009AA01A349, the Fundamental Research Funds for the Central Universities, National Natural Science Funds of China under Grant No: 11178020, 11175170 and the CAS Special Grant for Postgraduate Research, Innovation and Practice.

The authors are with the State Key Laboratory of Particle Detection and Electronics and Department of Modern Physics, University of Science and Technology of China, Hefei, Anhui 230026, China
(e-mail:wangjian@ustc.edu.cn)

Thus an expeditious error reconciliation protocol is needed. To achieve this, Field Programmable Gate Array (FPGA) is adopted in practical QKD systems, along with an error reconciliation protocol designed for the FPGA compatible device. The strength of FPGA is its parallel computing ability, large integrated RAM and easy bit-wise operation. Accordingly, the error reconciliation method we created considers these FPGA attributes. The proposed method is applied only to discrete variable QKD systems.

## II. DESCRIPTION OF THE RECONCILIATION AND ITS IMPLEMENTATION

### 2.1. Introduction to Cascade and Winnow

Before explaining Cascade[5], we first describe the error correction technique used in Cascade which is called BINARY. Consider two parties, say, Alice and Bob for simplicity, having strings A and B of length N, respectively. B is a little different with A where the discrepancy is estimated by the error rate p. BINARY works as follows. When Alice's and Bob's strings A and B have odd number of errors, Alice sends Bob the parity of the first half of her string. A comparison between this parity and the parity of the first half of Bob's is done to determine whether an odd number of errors exist in the first half, or in the second half, of Bob's string. This process is repeatedly applied to the half determined by the comparison until one bit is left, the erroneous bit. Finally, this error can be corrected.

Cascade proceeds in several passes. The number of passes is determined by Alice and Bob before execution depending on the parameter p. Let $A=A_1,\ldots,A_N$ and $B=B_1,\ldots,B_N$ (with $A_i, B_i \in \{0,1\}$) be Alice's and Bob's strings, respectively. In pass m, Alice and Bob choose $k_m$ and divide their strings into blocks of $k_m$ bits. The bits whose position is in $K^m_v = \{q \mid (v-1)k_m < q \leq vk_m\}$ form block v in pass m. Alice sends the parities of all her blocks to Bob. Using BINARY, Bob corrects an error in each block whose parity differs from that of Alice's corresponding block. At each pass $i>1$, Alice and Bob choose $k_i$ and a random function $f_i:[1..N]\rightarrow[1..\lceil N/k_i \rceil]$. The bits whose position in $K^i_j = \{q \mid f_i(q)=j\}$ form block j in pass i. Alice sends Bob the parity of the block $K^i_j$, $a_j = \sum_{q \in K^i_j} A_q$ for each $1 \leq j \leq \lceil N/k_i \rceil$. Bob computes his $b_j$'s in the same way and compares them with the $a_j$'s. For each $b_j \neq a_j$, Alice and Bob execute BINARY on the block defined by $K^i_j$. Bob will find $q \in K^i_j$ such that $B_q \neq A_q$ and correct it. All the blocks $K^u_v$ for $1 \leq u < i$ such that $q \in K^u_v$ will then have an odd number of errors. Let **K** be the set of these blocks. Alice and Bob can now choose the smallest block in **K** and use BINARY to find another error. Let w be the position of this error in strings A and B. After correcting $B_w$, Bob can determine set **B** formed by the blocks containing $B_w$ from each pass from 1 to pass i. He can also determine the set **J** of blocks with an odd number of errors by computing $\mathbf{J}=(\mathbf{B} \cup \mathbf{K})\setminus(\mathbf{B} \cap \mathbf{K})$. If $\mathbf{J} \neq \emptyset$ then Bob finds another pair of errors in the same way. This process is repeated until there are no more blocks with an odd number of errors, at which point pass i ends.

For security considerations, the information exchanged between the two sides should usually be eliminated after each pass of the reconciliation process. The information is composed of 1 bit for each parity check and $\log_2(k_i)$ bits for each execution of BINARY. However, the innovative aspect of Cascade is that it does not drop any bits during the entire reconciliation process. This technique improves the computational speed. A possible emergence of an error in a given pass can indicate errors in previous passes where the corresponding blocks contain the same error bit. Hence one error bit could reveal others. Although it has a high efficiency, Cascade is limited by its requirement for a large number of interactions. This causes great communication burden to QKD systems, especially in a public network environment.

Another widely used protocol is Winnow[9] in which Alice and Bob divide their key string into blocks according to an initial length typically determined by the parameter p. One side compares their parities and if they differ, a Hamming code is used to correct errors. Hamming code can fix a discrepancy in the block pair that contains only one error bit. Unlike BINARY, Winnow has the benefit of costing only one interaction which results in fewer communication exchanges and a faster process. This is particularly beneficial in situations where the classic channel has a high latency.

### 2.2. An error reconciliation based on FPGA

The Winnow protocol introduces the idea of correcting errors using Hamming code, a linear block code, to decrease interaction times. Building on this idea, we develop a reconciliation protocol well-suited to be implemented on an FPGA. The proposed protocol is as follows:

1. Initial length of block $n_0$ is determined by the bit error rate p between Alice and Bob, so the error rate must be evaluated first. In the reference introducing Cascade[5], the authors find optimum length $n_0$ is {73, 14, 7, 5} with the corresponding error rate p of {0.01, 0.05, 0.10, 0.15}. The observation of the relationship between the two parameters implies that $n_0*p$ is located between 0.7 and 0.8. Since the protocol is especially designed to work on an FPGA, the length being a power of 2 can simplify the entire error correction process. So we let $n_0$ be the largest value satisfying both $n_0*p<=0.8$ and a power of 2, but it is set to never less than 8. For example, if the error rate p is {0.01, 0.05, 0.10, 0.15}, $n_0$ is set as {64, 16, 8, 8}.
2. During the ith iteration, the full key string is divided into blocks of length $n_i$. Alice compares the parity of each block after receiving that of the corresponding block from Bob. If the parities for all blocks are identical or $n_i$ reaches $\lceil N/2 \rceil$, where N is the length of the key string, then it goes to step 5; otherwise to step 3.
3. For the blocks having different parity, a Hamming code is used to correct errors. Only blocks that contain one error can be corrected. For those holding more than one error, new errors may be introduced.

4. Double the length of block, that is $n_i=n_{i-1}*2$, permute the whole key string and revisit step 2.
5. Compare the 64-bit long cyclic redundancy check(CRC) code of the key strings of Alice and Bob. If they are equal, the key string is retained after eliminating the leaked information which is exchanged during the reconciliation. If they are not equal the key string is abandoned.

The permutation process in step 4, which requires a large amount of pseudo random numbers, heavily influences the performance of the protocol. A linear feedback shift register (LFSR) is widely used in FPGA design because of its simplicity and efficiency. An LFSR is a shift register whose input bit is a linear function of its previous state. The input bit is driven by the exclusive-or (XOR) of certain bits of the shift register. Because pseudo random numbers (PRNs) generated by the LFSR have a long cycle period, all the bits in a key string can be visited if the initial value of the LFSR, referred to as the seed, has a width equal to that of the length of the key string. However, just how to use these PRNs to develop an efficient permutation method is a difficult problem. A simple permutation scheme is designed as follows. Let $(a_0,...a_{N-1})$ and $(b_0,...b_{N-1})$ be two number sequences independently generated by two LFSRs using two different seeds. Let $a_i$ or $b_i$ be the position of two bits in the key string that are to be exchanged as i runs from 0 to N-1. The effectiveness of this permutation method will be shown in Section 3.2.1

### 2.3. Composition of a reconciliation module in the FPGA design

The FPGA design, following the description of the protocol, consists of four main modules: the interface module, the parity comparison module, the Hamming code module, and the permutation module. Fig. 2 shows the relationship among these modules.

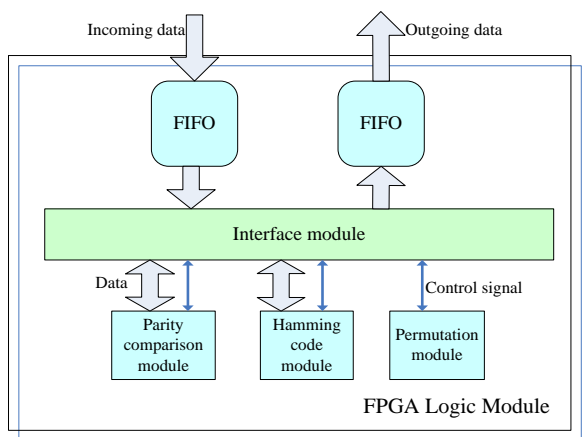

Fig.2 FPGA design of the reconciliation

1. Interface module

Interface module is designed to receive and send information exchanged during the whole reconciliation process. Each side requires only two First In First Outs(FIFOs): one for incoming data and one for outgoing data. This type of data exchange method conserves memory resources. Note that the parity comparison module, the Hamming code module and the permutation module are not organized in a pipeline fashion, because after each pass during the reconciliation, a permutation must be done prior to the next pass. A data bus and control module was developed to coordinate the use of the FIFOs among different modules. All modules connect to the data bus, while at any given time, and as determined by the control module, only one module serves as the data source.

Fig.3 illustrates the structure of the interface module. When the FIFOs are not full, each side is either performing parity comparison or Hamming error coding. Neither side is idle except for the case of excessive latency in the local net. When this occurs one side is prompted into a wait state.

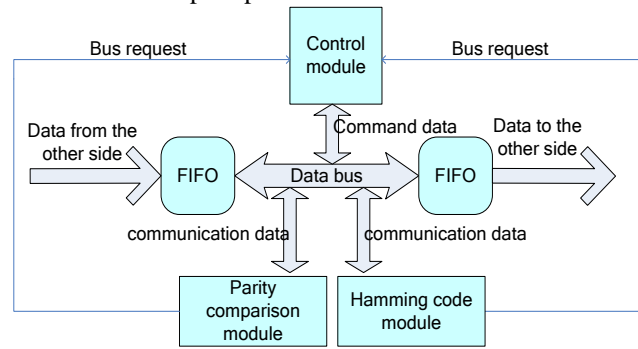

Fig.3 Composition of the interface module

2. Parity comparison module

Parity comparison module is responsible for calculating and comparing the parities of both sides. It also records the numbers of blocks which have different parities, and transfers the results to the Hamming code module. In the implementation of the hardware, parity computation is easily achieved using the single data operator '^' as provided by the hardware description language Verilog. For example, if the parity of a vector $x=(x_1,...,x_n)$ is desired, it can be expressed as parity(x)=^( $x_1,...,x_n$) in Verilog and computed in just one clock cycle. To achieve the same result using PC software however, requires n clock cycles since PC instructions lack a similar operator and the computation must be done in serial fashion.

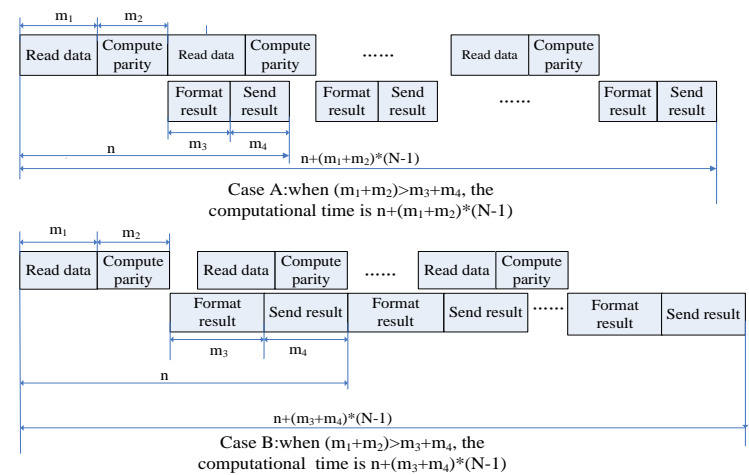

Fig.4 Pipeline method used in the parity comparison module

Another important advantage of FPGA is the ability to employ the widely used pipeline method. There are four steps to finish the parity comparison which separately require $m_1, m_2, m_3, m_4$ clock cycles as depicted in Fig.4. The pipeline method can be applied in either case A or case B depending on the specific distribution of the clock cycles required by each step. The number of the clock cycles on PCs is typically n*N, where $n=m_1+m_2+m_3+m_4$ which is usually called the latency and N is the length of the key string. As shown in Fig.4, if the pipeline method is used in the FPGA, approximately a fraction of $(m_3+m_4)/n$ in case A, and $(m_1+m_2)/n$ in case B, of the total clock cycles can be saved. The pipeline method is also used similarly in both the Hamming code module and the permutation module.

3. Hamming code module

Hamming code is based on a matrix whose size is $r*(2^r-1)$ and r is the number of rows in the matrix. Fig.5 is an example of 3*7 hamming matrix where r=3. Its element can be denoted as $h_{i,j}=j/2^{(i-1)}$ mod 2, where i is the row number and j is the column number. Since these elements can easily be obtained by the left-shift operation, the matrix need not be stored in the FPGA, but can instead be generated in real time whenever needed. This conserves memory resources. Another advantage for the FPGA is that the binary multiplication can be achieved with the XOR operation and thus only requires one clock cycle. For example, ( a*b mod 2)=^[$(a_1,\cdots,a_n)$ & $(b_1,\cdots,b_n)$]. For software running on a PC, this calculation would require n instruction cycles due to the lack of the single data operator "^" similarly as in the parity comparison module.

$$h^{(3)} = \begin{bmatrix} 1 & 0 & 1 & 0 & 1 & 0 & 1 \\ 0 & 1 & 1 & 0 & 0 & 1 & 1 \\ 0 & 0 & 0 & 1 & 1 & 1 & 1 \end{bmatrix}$$

Fig. 5 Hamming matrix when r=3

4. Permutation module

Permutation module requires good pseudo random numbers with a long cycle period. Two sequences of LFSR-based pseudo random numbers are adopted in the permutation module as introduced in section 2.2 above.

We refer to the combination of the four main modules described above as a reconciliation module. Each reconciliation module handles a length of 64Kbits of the key string. Several reconciliation modules may be combined on a single FPGA by considering the resources of the specific FPGA. In our test in Section 3, eight independent reconciliation modules are arranged in the FPGA resulting in the ability to handle 512Kbits of the key string.

III. EXPERIMENT SYSTEM AND RESULTS

3.1 A real-time QKD system

The reconciliation protocol is implemented on an FPGA using Verilog and is integrated in an actual QKD setup in order to test its performance. The QKD setup is constructed based on decoy encoding BB84[14]-[16] in order to ensure security and also to obtain a longer transmission distance. The qubit transmission rate is 20MHz. According to our experiment environment, the average photon number is set to 0.6:0.2:0[17],[18] which represents signal state, decoy state and vacuum state respectively, and the corresponding frequency ratio is 6:1:1. The QKD system used in the test is depicted in Fig.6. There are three main steps: key sifting, error reconciliation, and privacy amplification. The classic communication channel is composed of USB connection between the FPGA and the PC, and an internet connection between PCs of Alice and Bob. This channel can satisfy the QKD systems with a photon emitting rate of 100MHz.

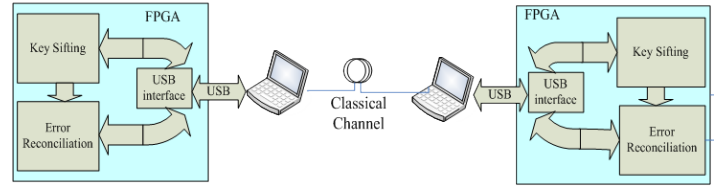

Fig.6 A Real-time QKD system

3.2 Results

3.2.1 Efficiency of the proposed permutation

Due to the parity check, most blocks hold an even number of errors after one pass. Errors from the same block must be separated into different blocks before the next pass so that they can be detected and corrected. Let $n_i$ be the length of the block during the ith pass and $A_{p,n_i}$ and $A_{q,n_i}$ be two bits residing in the same block, where p and q are two position subscripts of the key string. Permutation function is denoted by g. The key string is divided into blocks $K_{n_i,j}$ where j is the sequence number of the blocks, $0 \leq j \leq \lceil N/n_i \rceil$. We define a position function

$$P(A_{p,n_i})=k \text{ if } A_{p,n_i} \in K_{n_i,k}.$$

If two bits share the same block sequence number in two continuous passes, that is

$$P(A_{p,n_{i-1}})=P(A_{q,n_{i-1}}) \text{ and } P(A_{m,n_i})=P(A_{n,n_i})$$

with

$$A_{m,n_i}=g(A_{p,n_{i-1}}) \text{ and } A_{n,n_i}=g(A_{q,n_{i-1}}),$$

where m and n are the two corresponding position subscripts of the key string after permutation, we say the two bits are neighboring bits. A distance function $d(A_{p,n_{i-1}}; A_{q,n_{i-1}})$ can be defined as $d(A_{p,n_{i-1}}; A_{q,n_{i-1}})=0$ if $A_{p,n_{i-1}}$ and $A_{q,n_{i-1}}$ are neighboring bits, $d(A_{p,n_{i-1}}; A_{q,n_{i-1}})=1$ if they are not.

Let **B** be the set of bits who share the same block sequence number with $A_{p,n_{i-1}}$. **B** has $n_{i-1}-1$ elements represented by $B_{q,n_{i-1}}$. We define a function revealing the degree of the separation of a bit with the subscript p from other bits in the same block, $D_p$ as

$$D_p = \sum_{n_{i-1}-1} d(A_{p,n_{i-1}}; B_{q,n_{i-1}})/(n_{i-1}-1).$$

One would hope that $D_p=1$, indicating that there are no neighboring bit pairs and the reconciliation process would accelerate. In the worst case $D_p=0$, implying the permutation cannot separate an erroneous bit from other erroneous bits in the same block, and the erroneous bit would remain undetected in the subsequent pass.

To gain a global perspective, we define

$$D_{tot} = \sum_{p=0}^{N-1} D_p / N,$$

where N is the length of the key string. Similar to $D_p$, $D_{tot}$ indicates the degree of the separation on average, and one desires $D_{tot}$ to be near 1, the larger the better. In our test, the seed of the first LFSR was set to 5, whereas the seed of the second LFSR varied from 1 to N=64K. We measured $D_{tot}$ for the varying seeds of the second LFSR using $n_i=16$. From Fig.7, one can find $D_{tot}$ preserves high value greater than 0.99 at most points of the seed except for several ones. When we changed the seed of the first LFSR to another fixed value and repeated the above test, we got the similar results. In a specific permutation method, the seed of the first LFSR can be set to a constant, while the seed of the second LFSR has a wide selection range. One needs only to skip the several points where $D_{tot}$ deteriorates severely.

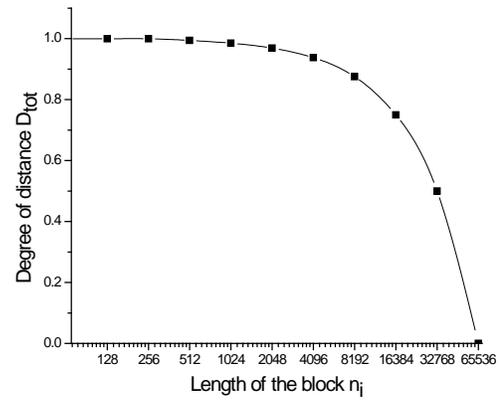

Fig.8 How $D_{tot}$ deteriorates when $n_i$ increases

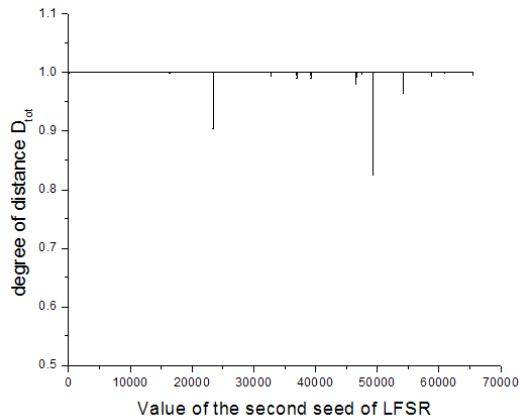

Fig.7 $D_{tot}$ with different value of the seed of the second LFSR

Fig.8 shows $D_{tot}$ as a function of $n_i$ where we set the seed of the first LFSR to 5 and the seed of the second LFSR to 78. $D_{tot}$ deteriorates for increasing $n_i$ and decreases sharply for $n_i \geq 8192$. When $n_i \geq 16384$, $D_{tot} \leq 0.75$ indicating the specialized permutation method is not effective for use in the reconciliation process. As such, one would desire that in most cases the reconciliation would be completed before $n_i$ reaches 16384. In fact, the reconciliation usually succeeds prior to $n_i=8192$.

For the purpose of comparison, we consider another simple permutation method which uses only one LFSR but with a much less efficient $D_{tot}$. The method acts as follows. Let $(c_0, \ldots, c_{N-1})$ be the number sequences generated by an LFSR, and the LFSR seed width be equal to that of the length of the key string. Through the sequence number q from 0 to $\lceil (N-1)/2 \rceil$, we exchange the position of every pair of bits represented by $c_{2q}$ and $c_{2q+1}$. This method also guarantees every bit of the key string can be visited and permuted. However, $D_{tot}$ turns out to be 0.76 using $n_i=16$, indicating this is not a good permutation method.

*3.2.2 Efficiency and process time of the proposed reconciliation*

There are two important parameters in evaluating the performance of the proposed reconciliation protocol: the total leaked information I(A;B) and total process time t. I(A;B) is the total bits transmitted between the two sides on the channel which may be exposed to an eavesdropper. Here A stands for Alice and B stands for Bob which are the two communication sides during the reconciliation process.

We first consider the leaked information I(A;B). The theoretic lower limit of the amount of leaked information, also called the Shannon entropy, is

$$h(p) = (-p\log(p) - (1-p)\log(1-p)),$$

where p is error rate. In addition, the reconciliation efficiency is defined as

$$f = I(A;B)/(N \times h(p)),$$

where N is the length of the key string. A smaller f means a better reconciliation protocol and a protocol with f=1 is considered optimal. Fig.9 shows the efficiency as a function of error rate where we set N to 64Kbits using a single reconciliation module.

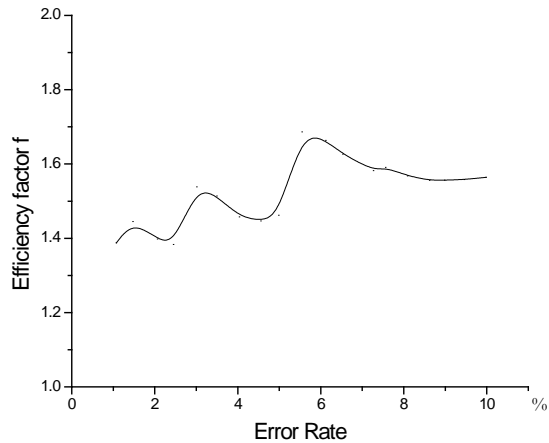

Fig.9 Efficiency f for different error rate p

The somewhat vibrative behavior of f may be explained by our choice for the initial block length as discussed in Section 2.2. In particular, for a certain range of p, $n_0$ takes on the same value, that is to say $n_0$ is not a continuous parameter as a function of p. When $n_0$ is {64, 32, 16, 8}, p should be the following corresponding minimal value {0.0125, 0.025, 0.05, 0.1} and this is consistent with the local minimums in Fig.9. We conclude that $n_0*p=0.8$ is a satisfactory choice of the initial length.

We now consider the total process time t for the reconciliation protocol. In this case we use eight reconciliation modules and set N to 512Kbits. Fig. 10 shows the results where t is measured by ignoring the latency and network transmission time. That is, the time shown is purely that of the required computation needed by the reconciliation modules. When integrated in real QKD systems as was done in section 3.1, there is typically an additional increase of around 200 ms to the overall processing time.

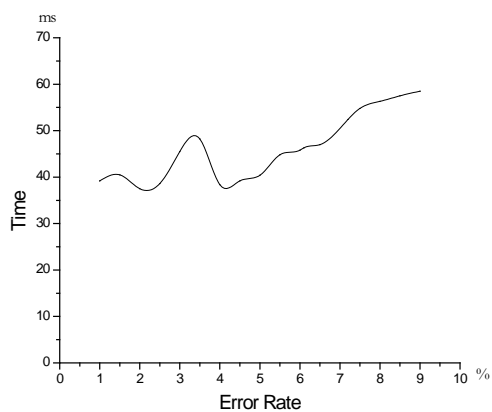

Fig.10 Time t for different error rate p

For a real QKD system, the error rate is generally below 4%. According to Fig.10 then, error reconciliation can be done in less than 50 ms. This implies we can adapt this reconciliation method to a sifted key rate of at least 10.5Mbits/s (i.e. 512Kbits/0.05s). This has an improvement compared with other reported reconciliation protocols[19]-[21] among which the maximal speed is 5.5Mbits/s using Cascade running on PC[21]. In addition, for comparison the same reconciliation on a PC with dual-core 2 Ghz processors and pure computational time of 360 ms results in a speed 1.5 Mbits/s.

Finally, we should note that hardware environment must be taken into account when considering the performance of a protocol. This work used the EP3C120F780C7 FPGA manufactured by Altera. This is a rather low-end product in the FPGA family. The operating clock is set to 100 MHz. The total resource cost is 30182/119088(25%) for the logic element and 1093632/3981312(28%) for the RAM - both relatively low. The advantages of the FPGA's parallelism provide for the integration of more reconciliation modules in order to improve performance without additional considerations. In other words, we consider such systems highly upgradeable.

## IV. CONCLUSION

We have proposed an expeditious FPGA-based error reconciliation method implementable in practical QKD systems. Benefits of the method include low hardware and time requirements, and the ability to easily upgrade such systems for further performance enhancement. This work helps solve the common problem of error reconciliation acting as a bottleneck in QKD systems.

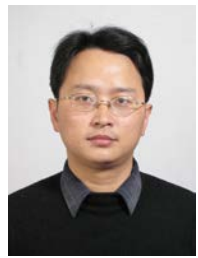
Jian Wang received Ph.D degree from University of Science and Technology of China(USTC) in 2003.
He is an associate professor in the Depart of Modern Physics at University and State Key Laboratory of Particle Detection and Electronics at USTC, P.R.China. His research interests include quantum key distribution systems, physical electronics, real-time processing technology.

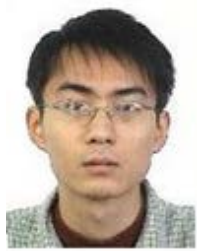
Ke Cui is a Ph.D student in the State Key Laboratory of Particle Detection and Electronics and Department of Modern Physics at University of Science and Technology of China (USTC). He received B.E. degree from USTC in 2008. His research interests include quantum key distribution systems, physical electronics, real-time processing technology.

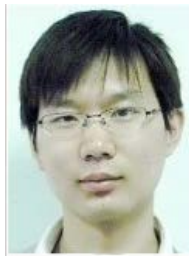
Hong-fei Zhang is a Ph.D student in the State Key Laboratory of Particle Detection and Electronics and Department of Modern Physics at University of Science and Technology of China (USTC). He received B.E. degree from USTC in 2009. His research interests include quantum key distribution systems, physical electronics, real-time processing technology.

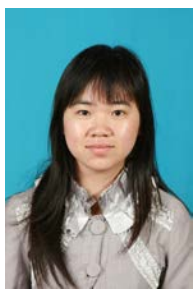
Chun-li Luo is a M.D student in the State Key Laboratory of Particle Detection and Electronics and Department of Modern Physics at University of Science and Technology of China (USTC). He received B.E. degree from USTC in 2010. His research interests include quantum key distribution systems, physical electronics, real-time processing technology.

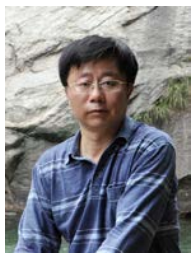
Ge Jin received Ph.D degree from University of Science and Technology of China(USTC) in 1989.
He is a professor in the Depart of Modern Physics at University and State Key Laboratory of Particle Detection and Electronics at USTC, P.R.China. His research interests include quantum key distribution systems, physical electronics, real-time processing technology.

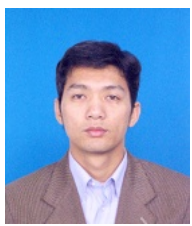
Teng-yun Chen received Ph.D degree from University of Science and Technology of China(USTC) in 2006
He is an associate professor at University of of Science and Technology of China, P.R. China. His research interest is quantum key distribution systems